\documentclass[a4paper, 10pt]{article}

\usepackage[a4paper, includeall, vmargin=1cm]{geometry}
\usepackage[LGR,LGR,T1]{fontenc}
\usepackage{amsmath,amssymb,array}
\usepackage{amssymb}
\usepackage{textcomp}
\usepackage{amstext}
\usepackage{graphicx}
\usepackage{setspace}
\usepackage{esint}
\usepackage{caption}
\usepackage[]{latexsym}
\usepackage{epsfig}
\usepackage{amssymb}
\usepackage{makeidx}
\usepackage{longtable} 
\usepackage{eucal}
\usepackage{subfigure}
\usepackage{latexsym}
\usepackage{fancyhdr}
\usepackage{authblk}
\usepackage{amsfonts}
\usepackage{floatflt}
\usepackage[latin1]{inputenc}
\usepackage{bbold}
\usepackage{mathrsfs}
\usepackage{sectsty}
\usepackage{lipsum}
\usepackage{amscd}
\usepackage{dsfont}
\usepackage{hyperref}
\usepackage{amsthm}
\hypersetup{
pdftitle={},%
pdfauthor={},%
pdfsubject={},%
pdfkeywords={},%
colorlinks=true,%
linkcolor=blue,%
linktocpage=true,%
hyperfootnotes=true,%
pageanchor=true
}
\input xy
\xyoption{all}

\begin{document}

\title{Full Characterization of the First 1 Inch Industrial Prototype of a New Concept Photodetector}
\author[a,b]{G. Barbarino}
\author[a,b]{F.C.T. Barbato}
\author[b]{C. M. Mollo}
\author[a,b]{E. Nocerino}
\author[a,b]{D. Vivolo}

\affil[a]{\textit{\footnotesize University of Naples "Federico II", Department of Physics "Ettore Pancini", Via Cintia, Naples, Italy}}
\affil[b]{\textit{\footnotesize INFN-Section of Naples, Italy}}

\maketitle

\begin{abstract}
\noindent
The VSiPMT (Vacuum Silicon PhotoMultiplier Tube) is an original design for an innovative light detector we proposed with the aim to create new scientific instrumentation for industrial applications and physics research. The idea behind this device is to replace the classical dynode chain of a photomultiplier tube with a silicon photomultiplier, the latter acting as an electron detector and amplifier.
\\
The VSiPMT offers very attractive features and unprecedented performance, definitely superior to every other photodetector with comparable sensitive surface, such as: negligible power cosumption, excellent photon counting, easy low-voltage-based stabilization and very good time performance. 
\\
After the feasibility test of the idea, Hamamatsu Photonics realized for our research group two VSiPMT industrial prototypes, that have been fully characterized. 
\\
The results of the full characterization of the 1-inch industrial prototype are presented in this work.
\end{abstract}

\section{Introduction}
\begin{figure}[h!]
\centering
\includegraphics[scale=0.4]{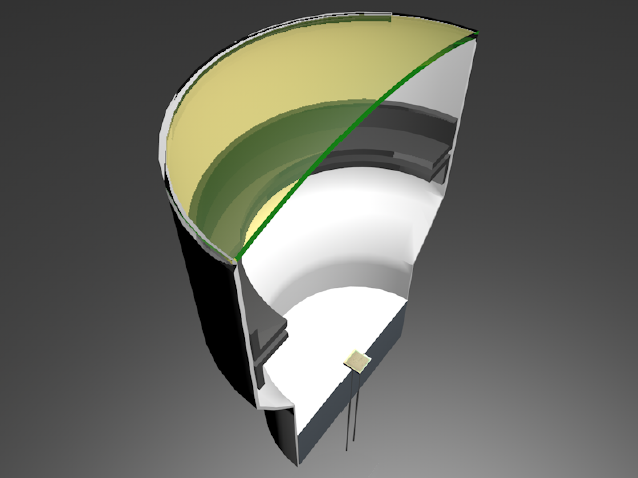}
\caption{A schematic cutaway of the VSiPMT showing the interior composition of the device. On the top there is the light entrance window, then a photocathode for the photons conversion into electrons. In the middle there is a focusing ring producing an electric field which accelerates and focuses the photoelectrons on the SiEM surface. Finally, on the bottom there is the SiEM acting in this configuration as an electron multiplier. Everything
is assembled into and hermetically sealed container.}
\label{vsip}
\end{figure}

Photon detection is a key factor to study many physical processes in several areas of fundamental physics research (i.e. particle and astroparticle physics, biomedicine, nuclear physics), industrial applications (i.e. environmental measurement equipment, image intensifier), medical equipment and biotechnology (i.e. PET, Radioimmunoassay and Enzyme immunoassay as luminescent, fluorescent, Chemiluminescent Immunoassay).
This impressive range of applications requires an equally impressive amount of detectors that can fulfill the requirements of each specific case. 
\\
In astroparticle physics experiments, the most currently used photodetectors are the PMTs (PhotoMultiplier Tubes), an almost one century old technology that hydes some limits. The worst features of PMTs are, indeed: the limitation in photon counting and the considerable power consumption.
\\
These limits pushed the progress of solid state photodetectors, such as the Silicon PhotoMultiplier (SiPM).
\\
Thanks to its different multiplication principle, this device boasts among its feature excellent photon counting capability, negligible power consumption and low costs. The SiPM shows a long lifetime, and is insensitive to magnetic fields. Nevertheless, the use of SiPMs in large area experiments is strongly limited by the small size (typically not exceeding $\sim$ 1 cm$^2$), mainly due to the internal noise.
\\
We invented a new high-gain, siliconbased photodetector, the Vacuum Silicon Photomultiplier Tube in order to exploit the SiPM performance in detectors with a larger sensitive surface. The basic idea, born in Naples in 2007 \cite{2}, is to match a special SiPM called SiEM (Silicon Electron Multiplier) with a large area photocathode. In this configuration the photocathode converts photons into photoelectrons and the SiEM provides to their multiplication, acting as an electron multiplier (figure \ref{vsip}).

\noindent
In 2012, after the proof of concept \cite{prof}, Hamamatsu Photonics realized for us two industrial prototypes: EB-MPPC050 (ZJ5025) and EB-MPPC100 (ZJ4991). 
Their characterization highlighted performance beyond the most optimistic expectation: excellent photon counting, low power consumption, easy stabilization and very fast response \cite{prof2}.
\\
Last summer Hamamatsu Photonics realized a 1-inch industrial prototype of VSiPMT: the EB-MPPC100 (XE2597) 1 INCH.

\section{The 1-Inch Industrial Prototype}
\begin{figure}[h!]
\centering
\includegraphics[scale=0.6]{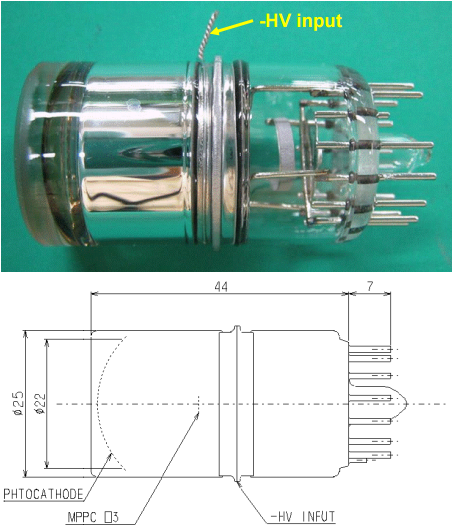}
\caption{Picture of the VSiPMT prototype (side view) together with the dimensional outline of the device in mm.}
\label{vsipmt1}
\end{figure}

The EB-MPPC100 (XE2597) is the first actually usable VSiPMT prototype (figure \ref{vsipmt1}). The absence of the dynode chain entails a visible consequence: the device is much more compact than a 1 inch PMT.

\noindent 
In this device a SiEM is mounted. To turn a SiPM into a SiEM two conditions are necesary:

\begin{itemize}
\item a p over n configuration of the junction for the Geiger avalanche trigger efficiency optimization,
\item the reduction of the SiO$_2$ passivation layer\footnote{SiPMs have a layer of silicon dioxide between the silicon wafers and the metal connections (typical thickness 1$\mu$m) in order to reduce any parasitic effects due to the interaction between the two materials.} thickness to facilitate the photoelectron penetration in the silicon bulk (from our simulations it has to be 200 $\div$ 300 nm \cite{proc1,proc2}).
\end{itemize}

\begin{figure}[h!]
\centering
\includegraphics[scale=0.75]{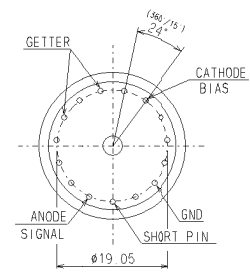}
\caption{Detail of the connenctions.}
\label{vsipmt2}
\end{figure}

\noindent
As in the first prototype, also in this case we have only three output connections: the voltage supply for the SiEM, the readout signal and the HV. In addition we have also a ground pin (figure \ref{vsipmt2}). 

\begin{figure}[h!]
\centering
\includegraphics[scale=0.6]{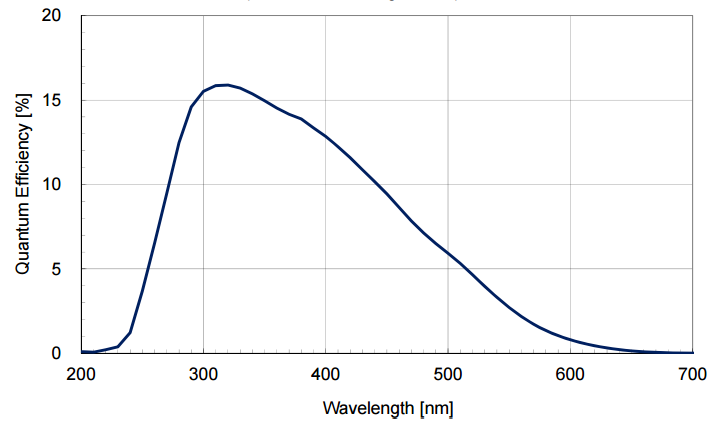}
\caption{Photocathode spectral response.}
\label{spe}
\end{figure}

\noindent
The device is composed by a 1-inch photocathode, a single focusing stage and a 3x3 mm$^2$ SiEM.
The bialkali photocathode has a quantum efficiency (QE) of $\sim$12$\%$ at 407nm (figure \ref{spe}). The QE is not optimized for this device, but it is sufficient to allow us to test the prototype.

\noindent
A single stage focusing system is present to drive the photoelectrons on the SiEM. It is composed by a focusing ring kept at the same potential with respect to the HV applied to the photocathode. 
\\
Characteristics and recommended work conditions are summarized below:

\begin{itemize}
\item SiEM area: 3 x 3 mm$^2$;
\item pixel: n.900, size 100 $\mu$m;
\item fill factor: 78$\%$;
\item optimized configuration:  p$^+$nn$^+$;
\item SiEM operation voltage: +71.5 V;
\item photocathode power supply: $-$1.9 kV.
\end{itemize}

\section{The Characterization}
The prototype underwent many tests summarized here below and discussed
in the following sections, in order to achieve a full characterization of:

\begin{itemize}
\item signal quality stability and photon counting capability;
\item detection efficiency;
\item gain;
\item photocathode omogeneity;
\item Transit Time Spread;
\item dark counts;
\item linearity and dynamic range.
\end{itemize}

\begin{figure}[h!]
\centering
\includegraphics[scale=0.4]{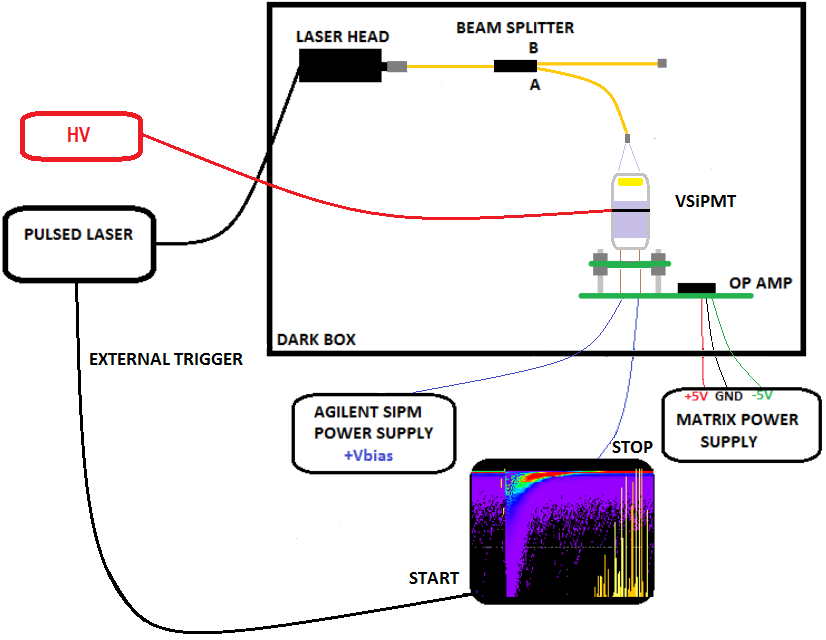}
\caption{Schematic view of our experimental set-up.}
\label{settts}
\end{figure}

\noindent
The experimental set up prepared for these measures is shown in figure \ref{settts}. The signal from the VSiPMT is amplificated by a high speed operational amplifier (Texas Instruments LMH6624) in inverting mode and the photocathode voltage is provided by a negative high voltage power supply (Caen WDT5533EXMAA). A luminous signal from a picosecond laser source (Hamamatsu M10306-04) with monochromatic emission at 407 nm is sent to the photocathode 
The data acquisition is provided by Digital Teledine LeCroy serial data analyzer mod. SDA 760Zi-A, 4 input channels, 6 GHz bandwidth with 40 GS/s sample rate.

\section{Signal Properties and Photon Counting Capability}
\begin{figure}[h!]
\centering
\includegraphics[scale=0.25]{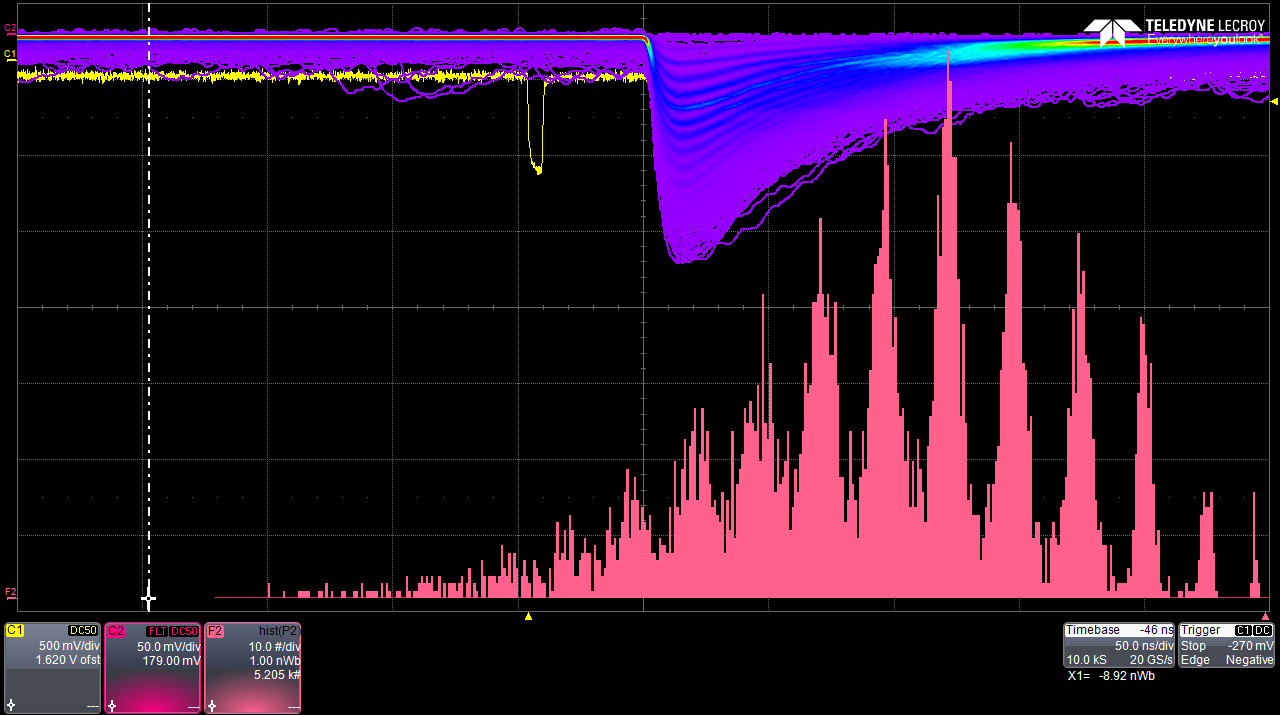}
\caption{Printscreen of the VSiPMT output signal in response to 90 incident photons and the corresponding charge histogram.}
\label{vsipmt}
\end{figure}

Every measurement is performed with the device output connected to a custom amplifier (x20 amplification), the voltage supply for the SiEM is V$_{bias}$ $=$ 71V and the HV for the photocathode is HV $=$ -1.9 kV.
In figure \ref{vsipmt} a superposition of EB-MPPC100 waveforms is kept. The charge spectrum shows the typical fingerplot shape, highlighting the excellent photon counting of the deivce. The data acquisition is triggered by the laser sync output, the responses for multiple triggers are overlayed. The measured signal pulse duration is (100 $\pm$ 1) ns with a rise time $\tau_{rise}$ $=$ (15 $\pm$ 1) ns and a fall time $\tau_{fall}$ $=$ (90 $\pm$ 1) ns.

\section{Gain}
\begin{figure}[h!]
\centering
\includegraphics[scale=0.43]{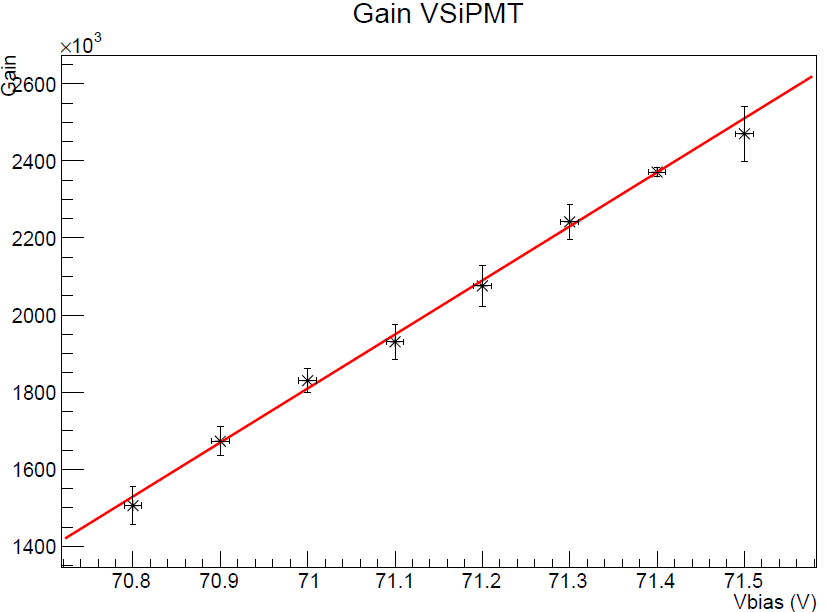}
\caption{Graph of gain versus V$_{bias}$ for the VSiPMT.}
\label{vsipg}
\end{figure}

In a VSiPMT the gain is obtained by the electrons crossing the Geiger region of the SiEM. A standard current signal is given for each fired cell, thus we expect the gain of the EB-MPPC100 to be that of the SiEM housed inside the tube.
\\
The gain of the SiEM can be defined as follows:

\begin{equation}
G = \frac{Q_{tot}}{e},
\end{equation}

\noindent
where Q$_{tot}$ is the total charge of the pulse generated by each pixel during the breakdown process and e is the charge of the electron. 
\\
A measure of the SiEM gain as a function of the applied bias voltage in the
range [70.8, 71.5] V has been performed, with the photocathode turned on. A linear trend is obtained as expected (figure \ref{vsipg}), with values ranging between 1.5 x 10$^6$ and 2.4 x 10$^6$. The same measure has been performed directly on the SiEM, keeping the HV off. We obtained both for the EB-MPPC100 and for the SiEM exactly the same values of the gain, so we plotted the gain only once in figure \ref{vsipg}.

\noindent
In a VSiPMT the gain is very stable with respect to variations in the low bias voltage, contrary to what happens in PMTs where the high voltage gain stabilization is much more difficult. Moreover, in a VSiPMT the gain does not depend on the linearity, contrary to what happens in PMTs, where gain is inversely proportional to linearity.

\newpage
\section{Noise}
\label{tabella}

The main contributions to the dark noise in a VSiPMT are  represented by dark pulses thermally generated (1 pe)  and crosstalk (2 pe) inside the SiEM. For this reason, measurements of the dark count rate at different SiEM bias voltages and different thresholds have been done.
\\
The bias voltage has been varied in the range [71, 76] V with steps of 0.1V, the photocathode is kept at its working voltage HV = -1.9 kV and the environmental temperature was 25$^\circ$C. In this configuration measurements of the dark noise contributions have been done using thresholds of 0.5, 1.5 and 2.5 pe.
As for the gain, the dark noise depends mainly on the SiEM housed in the tube, so dark noise measurements on the EB-MPPC100 are expected to be the same as those of the SiEM. The results are shown in the table below and are in excellent agreement with the expected dark count rate of the SiEM itself.

\begin{table}[h!]
\begin{center}
\begin{tabular}{|l|l|l|l|}
\hline
Tab 1&&&\\
\hline
$0.5 pe$ & threshold & (316, 9 $\pm$ 0.6) kcps & $V_{bias} = 71.5 V$\\
\hline
$1.5 pe$ & threshold & (200.0 $\pm$ 0.2) kcps & $V_{bias} = 71.5 V$\\
\hline
$2.5 pe$ & threshold & (120.21 $\pm$ 0.17) kcps & $V_{bias} = 71.5 V$\\
\hline
\end{tabular}
\label{vsipmtdark}
\end{center}
\end{table}

\noindent
Since the dark noise is mainly due to the SiEM, this feature is continuously improving with the progress of the standard SiPM production technology. In fact, the SiEM mounted on this VSiPMT prototype belongs to an old generation series, the table below shows the measured values of the dark rate for a SiPM belonging to a new generation series (S13360). 

\begin{table}[h!]
\begin{center}
\begin{tabular}{|l|l|l|l|}
\hline
Tab 2&&&\\
\hline
$0.5 pe$ & threshold & (183,8 $\pm$ 0.5) kcps & $V_{bias} = 54.8 V$\\
\hline
$1.5 pe$ & threshold & (3.3 $\pm$ 0.2) kcps & $V_{bias} = 54.8 V$\\
\hline
$2.5 pe$ & threshold & (42.7 $\pm$ 0.4) cps& $V_{bias} = 54.8 V$\\
\hline
\end{tabular}
\label{tab}
\end{center}
\end{table}

\noindent
The values of the dark count rate are referred to the SiEMs operating voltages.
The new devices differ from the older ones for the lower bias voltage required, so for their superior performance in terms of dark rate. Indeed, for the 1.5 pe treshold, the SiPM's dark rate is comparable with the PMT's dark rate. Moreover, the very latest SiPMs on the market exibit even better performance in terms of dark rate.
Therefore it is reasonable to expect a great improvement of this characteristic in the next generations of VSiPMT.

\section{Detection Efficiency: Operating Point}
\label{defect}

The Photon Detection Effciency (PDE) of the VSiPMT can be defined as
the product of four factors:

\begin{equation}
PDE = \epsilon_{PC} \cdot \epsilon_{focusing} \cdot \epsilon_{fill factor} \cdot \epsilon_{trigger}.
\label{equazione}
\end{equation}

\begin{figure}[h!]
\centering
\includegraphics[scale=0.6]{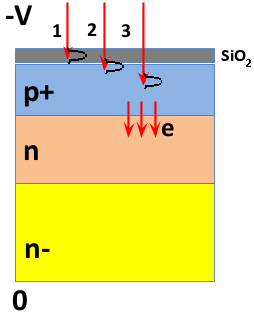}
\caption{Range straggling of the photoelectrons coming from the photocathode and impinging on the p-over-n SiEM inside the VSiPMT.}
\label{pe}
\end{figure}

\noindent
Where $\epsilon_{PC}$ is the photocathode's quantum efficiency, $\epsilon_{focusing}$ and $\epsilon_{fill factor}$ are the components of the geometrical efficiency which takes into account the focusing and the SiEM fill factor, $\epsilon_{trigger}$ is the trigger efficiency which depends on the HV. The latter is referred to the probability of an electron triggering the geiger avalanche (figure \ref{pe}).

\noindent
Referring to the figure, can be distinguished three cases for the triggering phenomenology:

\begin{itemize}
\item case 1: the range straggling of the electron is outside the p region ($\epsilon_{trigger}$ $=$ 0);
\item case 2: the range straggling of the electron is partially inside the p region (0 $<$ $\epsilon_{trigger}$ $<$ 1);
\item case 3: the range straggling of the electron is totally inside the p region ($\epsilon_{trigger}$ $=$ 1).
\end{itemize}

\noindent
The PDE is experimentally defined as:

\begin{equation}
PDE = \frac{N_{pe}}{N_{ph}},
\end{equation}

\noindent
where N$_{pe}$ is the number of fired cells and N$_{ph}$ is the number of photons per
laser pulse hitting the photocathode.
In the third case of trigger all the electrons have the right energy to enter into the depletion region. Therefore the PDE does not depend on the photoelectrons energy. This means that further stabilizations due to high voltage fluctuations are not necessary.
\\
For this prototype the SiEM fill factor is $\epsilon_{geom}$ = 78$\%$ and the photocathode efficiency at the wavelength under test is  $\epsilon_{PC}$ = 12$\%$, so in the third case for $\epsilon_{trigger}$ = 100$\%$ and considering an optimum focusing ($\epsilon_{focusing}$= 100 $\%$) the PDE is 9.4$\%$.
The PDE will udergo a remarkable improvement after the photocathode optimization, passing from $\epsilon_{PC}$ = 12$\%$ to $\epsilon_{PC}$ = 40$\%$, so it will be comparable with the standard PMT's PDE.

\section{Photocatode Scan X-Y}
\begin{figure}[h!]
\centering
\includegraphics[scale=0.6]{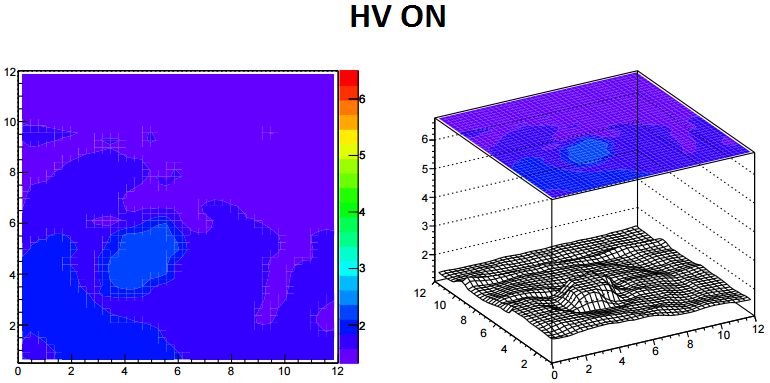}
\caption{Results of the x-y scan of the photocathode of the prototype with HV on.}
\label{xy}
\end{figure}

An x-y scan of the entire photocathode sensitive surface has been done to probe the PDE uniformity. For this measurement a micrometric x-y motorized pantograph has been used. Since the SiEM is also sensitive to photons, the scan has been done both with HV on and with HV off in order to estimate potential optical effects of the semitransparent photocathode on the SiEM.
\\
The scan performed with HV off showed a photocathode transparency effect of few $\%$  in correspondence of the SiEM. Neverthless, this is a negligible effect in all usual photodetection applications since the area affected by the phenomenon (9mm$^2$) is only 1.7\% of the total photocathode area (5cm$^2$), it will be even more neglectable in the next larger area devices (2 and 3 inches), and anyway it is an over-efficiency effect that does not affect the measurements. 
\\
Figure \ref{xy} shows xy efficiency scan when high voltage is supplied to the photocathode.

\noindent
This measurement has been performed by subtracting the over-efficiency due to the photocathode transparency to evaluate the homogeneity in the photocathode deposition, it results to be homogeneous within 90$\%$.

\section{Transit Time Spread}
\begin{figure}[h!]
\centering
\includegraphics[scale=0.6]{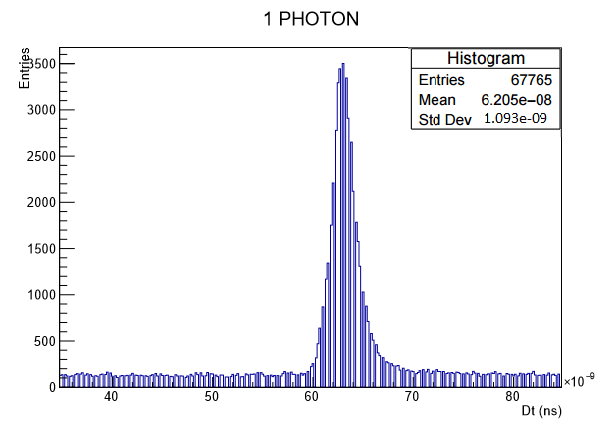}
\caption{Distribution of the time intervals between the single electron signals and the laser pulses made by means of the data analysis tool Root for 1 photon per pulse.}
\label{tts2}
\end{figure}

The TTS measurement has been
performed with the photocathode completely and homogenously illuminated by a one photon per pulse laser output. 
\\
We measured the time interval elapsed between a 0.5 pe signal and the sync out signal of the laser trigger (figure \ref{tts2}). We measured a TTS expressed in terms of FWHM equal to (2.576 $\pm$ 0.008) ns.

\noindent 
We evaluated the TT difference between a central path and a side path. To perform this measure we used a collimated laser beam with an average 100 photons per pulse close to the photocathode (figure \ref{tts3}):

\begin{figure}[h!]
\centering
\includegraphics[scale=0.4]{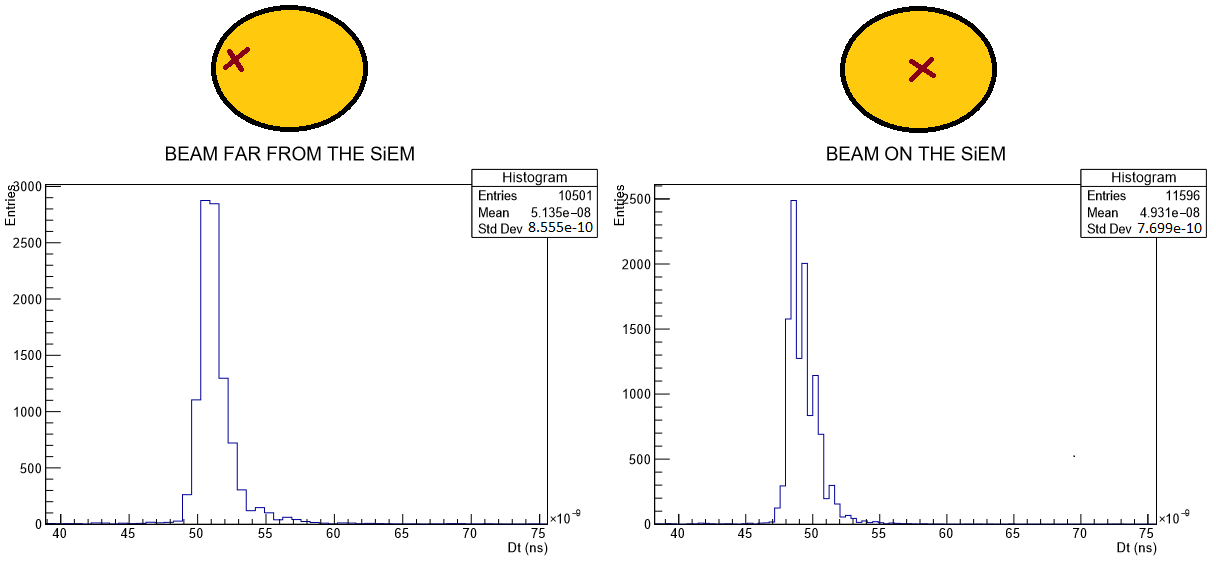}
\caption{Distribution of the time intervals between the single electron signals and the laser pulses made by means of the data analysis tool Root in the two cases. The little bulge on the right side of the two distributions is due to afterpulses.}
\label{tts3}
\end{figure}

\noindent
From this analysis it results:

\begin{equation}
TT_{diff} = (1.7966 \pm 0.0003) ns
\end{equation}

\noindent
The differences between the paths of photoelectrons which come from different points of the photocathode significatively contribute to the TTS. Indeed TTS and TT$_{diff}$ are of the same order of magnitude, by eliminating the difference between path contribution will be possible to have a TTS comparable with the PMT's TTS.

\section{Focusing}
\begin{figure}[h!]
\centering
\includegraphics[scale=0.95]{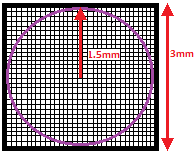}
\caption{Optimum focusing in our VSiPMT prototype.}
\label{focus}
\end{figure}

A correct focusing is a key point for the correct operation of a VSiPMT.
A too strong focusing produces a too much squeezed photoelectron beam. In this case the photoelectron spot intercepts only a fraction of the active surface of the SiEM, with a consequent reduction of the linearity. This reduction can be drastic, with the additional drawback that all the SiEM pixels not involved in the electron multiplication process are still dark count sources.
\\
However, a too weak focusing causes a photoelectron spot larger than
the SiEM surface. This means that a fraction of the photoelectron sistematically miss the SiEM surface, thus implying a reduction of the photon detection efficiency of the device.
\\
An optimal focusing condition is achieved if the photoelectron beam spot is perfectly inscribed in
the SiEM square target. In this case, no more than the 85$\%$ of the total number of cells of the SiEM can be fired. This is the configuration that exploits the maximum number of pixels leading to an optimized use of the device.
\\
To prove the focusing of our prototype, we targeted it with a laser beam at maximum power to let the SiEM reach the saturation and we measured the number of fired pixels changing the HV.
\\
The optimum solution for the focusing was 700 fired cells for a circular spot inscribed in the SiEM surface (figure \ref{focus}). 

\begin{figure}[h!]
\centering
\includegraphics[scale=0.9]{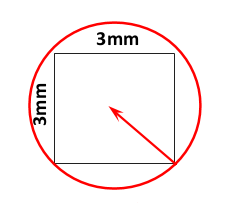}
\caption{Actual focusing in our VSiPMT prototype. The radius of photoelectron focusing area circumscribes the SiEM.}
\label{focus2}
\end{figure}

\noindent
In our case the SiEM is under-focused (figure \ref{focus2}), because we measured a number of fired cells equal to 900 $\pm$ 15, but, from our simulations on the SiEM placement \cite{proc1,proc2}, we know this is an easily optimizable feature. Further improvements will come with the use of round SiEMs, that let to us to eliminate the unused pixels and so the related dark rate. For example, referring to table 2 in section \ref{tabella}, the dark rate at 1.5 pe threshold would be 2.5 kcps instead of 3.3 kcps.

\section{Dynamic Range}
The dynamic range of a VSiPMT depends on the number of cells of the SiEM that are within the spot. The number of fired pixels (N$_{pe}$) with respect to the total number of pixels (N$_{cells}$) as well as to the incoming light  (N$_{ph}$) is: 
 
\begin{equation}
N_{pe}(N_{cells}, \lambda, V) = N_{cells} \cdot \left[1 - e^{\frac{-N_{ph} \cdot PDE (\lambda, V)}{N_{cells}}}\right].
\end{equation}

\begin{figure}[h!]
\centering
\includegraphics[scale=0.47]{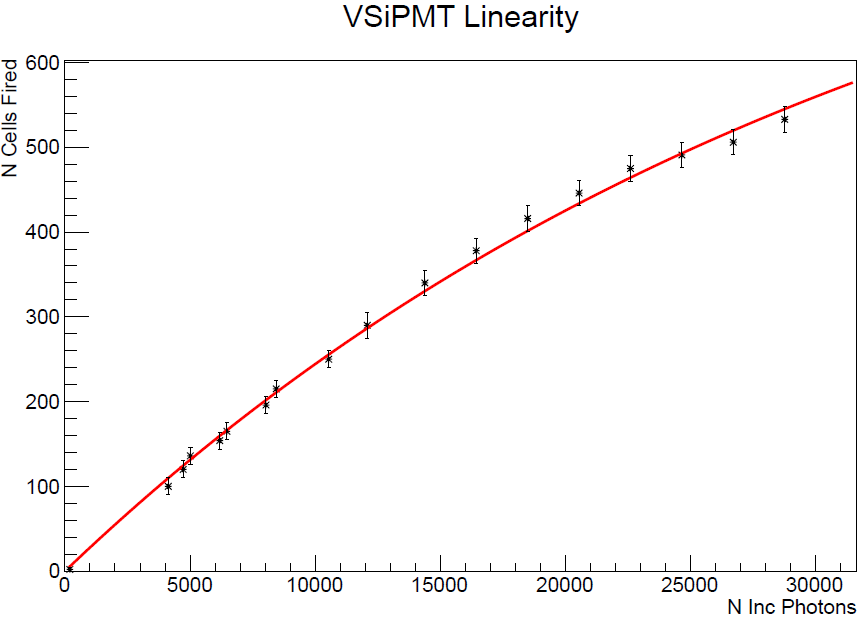}
\caption{Number of fired pixels as a function of the number of incident photons for the VSiPMT prototype.}
\label{vsiplin}
\end{figure}

\noindent
The graph in figure \ref{vsiplin} shows the number of fired pixels versus the number of incident photons (so the VSiPMT response).

\noindent
For $\sim$ 20000 incident photons the upper limit of the dynamic range has been reached, the response of the VSiPMT begins to deviate from linearity and the saturation occurs.
\\
From the fit we have:

\begin{equation}
N_{cells} = 945 \pm 100
\end{equation}

\noindent
The value of $N_{cells}$  is, within the error, equal to the total number of the SiEM's cells. 

\section{Conclusions}
In this work we presented the results of an extensive characterization of the latest usable VSiPMT prototype,  the EB-MPPC100 (XE2597) 1 INCH. This analysis provided first-rate results: the VSiPMT offers very attractive features and
unprecedented performances, definitely superior to every other photodetector of the same
sensitive surface; therefore it has the potentiality to fulfill the requirements of the next generation
of astroparticle physics experiments and not only. In fact, the outstanding performances and the features of the VSiPMT make its potential field of application go far beyond astroparticle physics; the VSiPMT can be considered a valid solution
in almost all the applications in which PMTs are adopted and in many applications in which the performances of a SiPM are required along with a larger sensitive surface. 
\\
The most outstanding achievements of this device are:
\begin{itemize}
\item negligible power cosumption,
\item excellent photon counting even in low light conditions,
\item easy low-voltage-based stability, 
\item high gain,
\item wide dynamic range, 
\item gain and linearity are independent.
\end{itemize}

\section{Acknowledgments}
The authors wish to thank Hamamatsu Photonics K. K. for the
realization of the prototype and the Italian Space Agency (ASI) for supporting this project.


\begin{thebibliography}{9}
\bibitem{2} G. Barbarino et al., A new high-gain vacuum photomultiplier based upon the amplification of a Geiger-mode p-n junction, Nucl. Instrum. Meth. A 594 (2008) 326-331.
\bibitem{prof} Barbarino G. et al., Proof of feasibility of the Vacuum Silicon PhotoMultiplier Tube (VSiPMT), Journal of Instrumentation, Volume 8, Issue 04, article id. P04021 (2013).
\bibitem{prof2} Barbarino G., Barbato F. C. T. et al., A new generation photodetector for astroparticle physics: The VSiPMT, Astroparticle Physics, Volume 67, p. 18-25, 2015.
\bibitem{proc1} F. C. T. Barbato, Proceeding RICH 2016, Bled.
\bibitem{proc2} C. M. Mollo, Proceeding NSS 2015, San Diego.
\bibitem{vac} Engstrom R. et al. Rca photomultiplier manual. RCA Electronic Components, Harrison, NJ, 1970.

\end{thebibliography}
\end{document}